*Original Article*

# Assessing the Impact of Image Super Resolution on White Blood Cell Classification Accuracy

Tatwadarshi P. Nagarhalli[1], Shruti S. Pawar[2], Soham A. Dahanukar[3], Uday Aswalekar[4], Ashwini M. Save[5], Sanket D. Patil[6]

[1,2,3]*Department of Artificial Intelligence and Data Science, Vidyavardhini's College of Engineering and Technology, Mumbai, Maharashtra, India.*

[4]*Department of Mechanical Engineering, Vidyavardhini's College of Engineering and Technology, Mumbai, Maharashtra, India.*

[5]*Computer Engineering Department, VIVA Institute of Technology, Mumbai, Maharashtra, India.*

[6]*Department of Computer Engineering, Vidyavardhini's College of Engineering and Technology, Mumbai, Maharashtra, India.*

[1]*Corresponding Author : tatwadarshipn@gmail.com*

***Abstract -*** *Accurately classifying white blood cells from microscopic images is essential to identify several illnesses and conditions in medical diagnostics. Many deep learning technologies are being employed to quickly and automatically classify images. However, most of the time, the resolution of these microscopic pictures is quite low, which might make it difficult to classify them correctly. Some picture improvement techniques, such as image super-resolution, are being utilized to improve the resolution of the photos to get around this issue. The suggested study uses large image dimension upscaling to investigate how picture-enhancing approaches affect classification performance. The study specifically looks at how deep learning models may be able to understand more complex visual information by capturing subtler morphological changes when image resolution is increased using cutting-edge techniques. The model may learn from standard and augmented data since the improved images are incorporated into the training process. This dual method seeks to comprehend the impact of image resolution on model performance and enhance classification accuracy. A well-known model for picture categorization is used to conduct extensive testing and thoroughly evaluate the effectiveness of this approach. This research intends to create more efficient image identification algorithms customized to a particular dataset of white blood cells by understanding the trade-offs between ordinary and enhanced images.*

## 1. Introduction

White Blood Cell (WBC) classification is important in medical diagnosis, especially concerning different haematological illnesses. To make appropriate treatment plans and identify illnesses, accurate classification is crucial. As stated in Gavas and Olpadkar [1]. Deep Convolutional Neural Networks (CNNs) have shown significant promise in automating the classification process by identifying complex patterns in massive datasets. Even with the significant advancements in CNNs, some difficulties remain, particularly when differentiating between closely related WBC types like neutrophils and eosinophils. These challenges often result in inaccurate classifications, which reduces overall diagnostic accuracy. The importance of resolving these issues was emphasized by T. Tamang, T. et al. [5] since minor misclassifications might have significant consequences in a clinical environment. To address these problems, researchers have investigated several strategies, including enhancing the quality and resolution of training pictures [11]. The shortcomings of conventional CNNs have mostly been overcome with image augmentation techniques, especially those targeted at raising the resolution of available datasets. Higher-resolution images can offer more detailed information, which helps models better identify similar WBC types, as Zhu, Z., Lei, Y., Qin, Y., Zhu, C., & Zhu, Y. (2023) [16] pointed out. However, these methods also bring additional challenges, like higher processing overhead and the requirement to modify the model to avoid overfitting [17]. Developing strong and dependable diagnostic tools requires understanding how to balance model performance with image resolution. Ren et al. (2020) [20]. It was pointed out that improved images might increase model accuracy, but their integration requires careful consideration to preserve the model's generalizability across various datasets and clinical circumstances. In order to progress in the field of medical picture categorization, this trade-off is essential. This work examines how sophisticated image resolution enhancement methods affect WBC classification. As mentioned by Sahaai et al. (2022) [29]. The goal is to have a system with good classification accuracy and aid in creating more effective diagnostic systems by combining these strategies with a CNN architecture explicitly designed for this objective, which can be achieved by understanding if image enhancement impacts image classification accuracy.

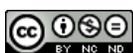

## 2. Literature Review

The system presented in the paper [3] addresses the challenge of accurately classifying WBCs, a crucial task in medical diagnostics. To improve feature representation, the authors suggested a unique method that combines a Spatial and Channel Attention Mechanism (SCAM) with pre-trained ResNet and DenseNet models. The study's classification accuracy demonstrated the technique's efficacy, which exceeded 90% using a WBC picture dataset. WBC classification is the domain of medical image categorization. This paper is critical to our research since it demonstrates how classification performance may be significantly enhanced by mixing several deep-learning models with attention mechanisms. In order to identify COVID-19, the work in [4] explores the classification of lung radiographs using a deep CNN architecture with ResNet-50. Among other respiratory conditions, the primary issue addressed is accurately identifying COVID-19 in chest X-rays. The scientists used lung radiography images to assist them in categorizing the images with an accuracy above 97%. Medical image analysis is the study's domain, emphasising COVID-19 detection. Although it has nothing to do with WBC classification, it shows how ResNet-50 may be effectively used for medical image classification, which raises the possibility that it could improve WBC image classification systems.

The proposed research in [9] assesses the usage of ResNet-50 for plant disease classification. With a dataset of plant disease photos, the authors obtained an accuracy of about 93%. Image analysis for agriculture is the domain. It shows the resilience of ResNet-50 in many classification tasks, which may be significant for enhancing WBC classification methods even though it focuses on plant diseases rather than WBCs. The proposed research in [11] uses the VGG 16, ResNet-50, and Inception V3 models to examine the categorization of brain tumours by transfer learning. The study tackles the issue of correctly diagnosing brain tumours via medical imaging. The scientists showed good classification accuracy using an image collection of brain tumours, with VGG16 demonstrating competitive performance. The field is medical imaging, more especially the identification of brain tumours. While the emphasis is on brain malignancies, ResNet-50's practical use in this situation may also enhance WBC categorization systems. The classification of WBCs using different hybrid CNN systems is the subject of this research [14]. Accurately identifying distinct WBC kinds in medical imaging has been highlighted as an issue. Using a dataset of WBC pictures, the authors could classify the photographs with excellent accuracy. Medical image analysis is the study's domain, and WBCs are its main emphasis. The method demonstrates how well hybrid CNN models classify WBCs, which is consistent with our investigation of cutting-edge methods to enhance WBC image analysis. Real-ESRGAN [17], a technique for improving image resolution with synthetic data, is presented in this study. Training models using exclusively synthetic datasets solves the problem of applying super-resolution techniques to real-world photos. The authors showed that Real-ESRGAN achieves noteworthy outcomes in boosting image details while also efficiently improving image quality and resolution. Generative Adversarial Networks (GANs) are one of the technologies employed in the picture super-resolution domain. This method is pertinent to our work since it offers insights into sophisticated image-enhancing approaches that may help improve WBC image categorization using high-resolution image inputs.

The paper [20] explores using GANs for real-world super-resolution tasks. Improving image resolution in real-world situations-where high-quality data is frequently unavailable-is the issue being tackled. The authors showed notable increases in image resolution by using GANs to enhance image clarity and detail. The study area is super-resolution images, emphasising using GANs in real-world scenarios. This study utilizes advanced image enhancement methods to categorize WBC, potentially improving the performance of classification models. A novel method for super-resolution [24] that improves image details and gives the user more control over the resolution process is presented in this study. The authors use cutting-edge generative models to tackle the problem of enhancing visual quality and clarity. The strategy, which uses deep learning and GANs, outperforms current techniques. This work is relevant to our research since it provides sophisticated image-enhancing techniques that may improve WBC classification by yielding sharper, more detailed images.

The study's goal in [27] is to apply regional CNN, YOLO v3, for detecting and classifying WBCs. The precise detection and categorization of WBCs in medical pictures is the issue that is being addressed. The authors reported successful classification results after using YOLO v3 to increase detection precision. The study highlights the effectiveness of CNN-based techniques in classifying WBCs, highlighting the potential of advanced neural networks to enhance WBC classification accuracy. The initial goal of this work [31] was to create a deep-learning model for automated categorization of WBCs. Despite initially aiming for accuracy, the study utilized deep learning techniques to automate the classification of WBCs, highlighting the importance of sophisticated neural network techniques. CNN image recognition algorithms are examined in this research [38], emphasising improving the precision of AI-driven picture analysis. The paper highlights problems with picture recognition and suggests employing cutting-edge CNN architectures to address them. It discusses how various CNN models may enhance classification efficacy on various datasets. This work offers insightful information on using CNNs for image recognition applications that is pertinent to our WBC classification study. We may improve the precision and efficiency of our WBC categorization system by implementing these cutting-edge methods. The research [42] examines the classification of leaf diseases in various crops using the VGG CNN and transfer learning. The work uses transfer learning techniques and pre-trained models to tackle the problem of accurate illness categorization. The authors show that classification accuracy is increased by optimizing the VGG model using datasets related to leaf disease. Combining CNNs and transfer learning yields insights for maximizing model performance (98.4%) for classification tasks, even if the focus is on agricultural photos. These insights may be used in our WBC classification project to improve accuracy.

Table 1. Performance metrics of deep learning algorithms for image classification

| Research Paper | Techniques Algorithms Used | Dataset Used | Accuracy |
|---|---|---|---|
| [3] | ResNet and DeneNet | WBC | > 90% |
| [2] | ResNet50 | COVID-19 | 97.0% |
| [9] | ResNet50 | Plant Disease Dataset | 93% |
| [27] | YOLO v3 | WBC | 96% |
| [38] | ResNet50 | Medical Dataset | 93.51% |
| [42] | VGG | Plant Disease Dataset | 98.4% |

Table 1 overviews many deep learning algorithms and their performance in diverse domains' picture classification tasks. It demonstrates the application of methods for classifying images using pre-trained models.

The table shows how these cutting-edge techniques have been used on various datasets, showcasing their wide range of applications and the gains in classification tasks. It shows how much deeper learning models have advanced our ability to classify images accurately in agricultural and medical settings.

## 3. Proposed System
In medical diagnostics, precise classification of WBCs is essential. The research needs a particular data set where all kinds of WBC images are classified into different groups. In this study, the paper must compare standard and enhanced images. The paper's results emphasize the result of making those changes.

*3.1. Overview of the Dataset*
The proposed work uses a standard dataset for categorising blood cells obtained from Kaggle. Four different types of WBCs are depicted in the images: neutrophils, lymphocytes, eosinophils, and monocytes.

Table 2. Description of the dataset

| Class /Set | Train | Test |
|---|---|---|
| Eosinophils | 2,497 | 623 |
| Lymphocytes | 2,483 | 620 |
| Monocytes | 2,478 | 620 |
| Neutrophils | 2,499 | 624 |
| **Total** | 9,957 | 2,487 |

Table 2 describes the four types of WBCs in the dataset: neutrophils, lymphocytes, monocytes, and eosinophils. There are 2,487 photos for testing and 9,957 images for training and validation, with 7968 and 1989 images, respectively, with almost equal representation of each class in both sets.

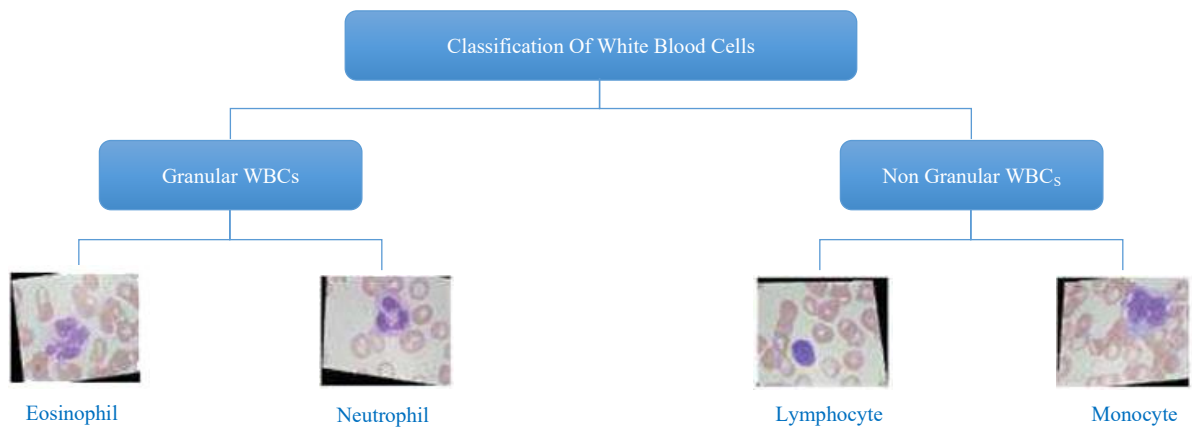

Fig. 1 Classification of WBC's

Figure 1 illustrates the four main kinds of WBCs, namely neutrophil, monocyte, lymphocyte, and eosinophil, along with their unique granular properties. Many disorders require accurate categorization of these cells to be monitored and diagnosed. In Figure 2, the Eosinophil picture is from the original dataset used to compare the datasets. The image of eosinophil is smaller in dimensions of 320 x 240 and resembles a minute.

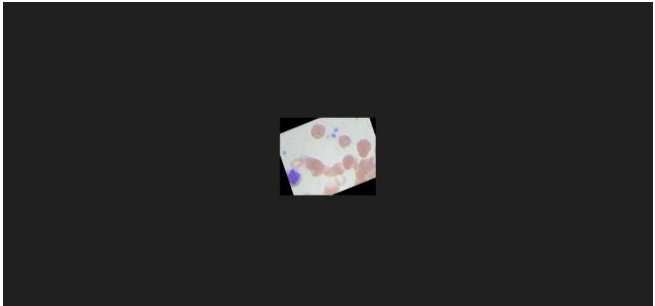

**Fig. 2 Size of the original image (Eosinophil) with magnification 65%**

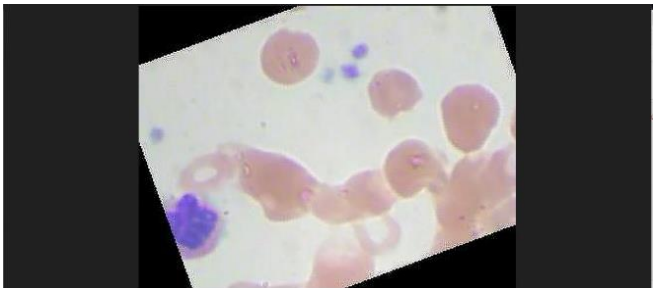

**Fig. 3 Size of enhanced image (Eosinophil) with magnification 65%**

In Figure 3, the Eosinophil picture is from the enhanced dataset used for comparing the datasets. The image of eosinophil is larger, in dimensions of 1280 x 960, and resembles a big one.

### 3.2. Model Used for Image Classification

A significant advancement in addressing the difficulties involved in training profound neural networks is ResNet or Residual Networks. Traditional deep networks have issues with vanishing gradients, where the gradients grow smaller and smaller as they go through the network, making it harder for the model to train well. In order to go around one or more convolutional layers, ResNet introduces residual blocks with skip connections [2].

These skip connections maintain the gradient's magnitude and guarantee efficient backpropagation by allowing the gradient to pass through the network directly. Thanks to this architectural breakthrough, networks with hundreds of layers may be trained without experiencing the degradation issues that deep networks usually face.

So, ResNet models perform better on many tasks, such as object identification and picture categorization. Since residual learning makes it possible to train even deeper networks-which can learn more complicated features and representations-the subject of deep learning has evolved substantially. This has opened the door for a wide range of applications in computer vision and other fields requiring a complex and nuanced interpretation of data [4]. An overview of this model can be found in Figure 4.

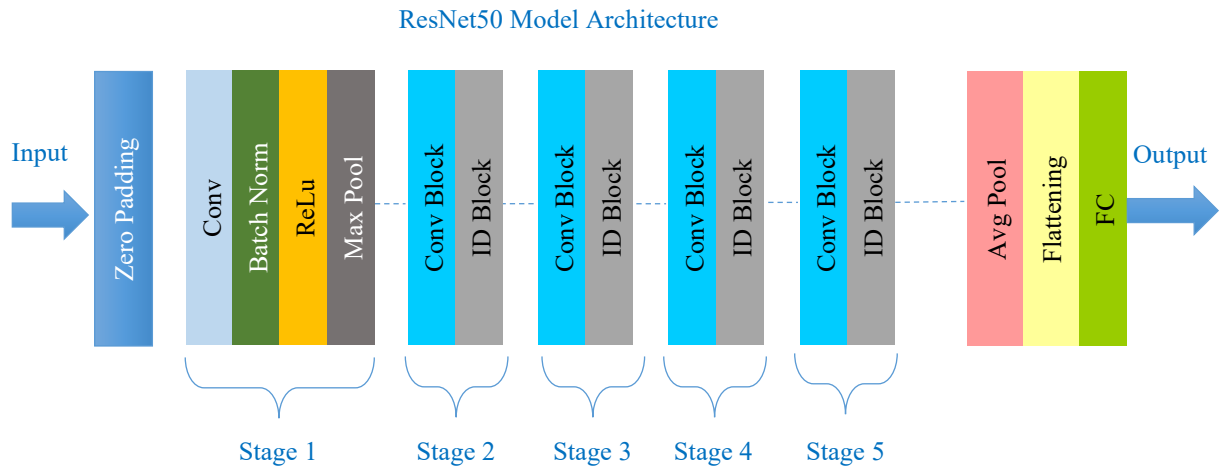

**Fig. 4 Overview of the layer-by-layer structure and flow of the ResNet-50 architecture**

The ResNet-50 deep CNN is designed with a ReLU activation function, batch normalization, max pooling, and an initial convolutional layer. The network's core consists of multiple convolutional and identification blocks, with a fully connected layer producing class predictions [7]. ResNet-50 uses residual learning to simplify deep network training, with skip connections preventing performance deterioration. The architecture stabilizes and speeds up the training process, ensuring gradient information retention for intricate patterns and features, making it suitable for image classification applications [9].

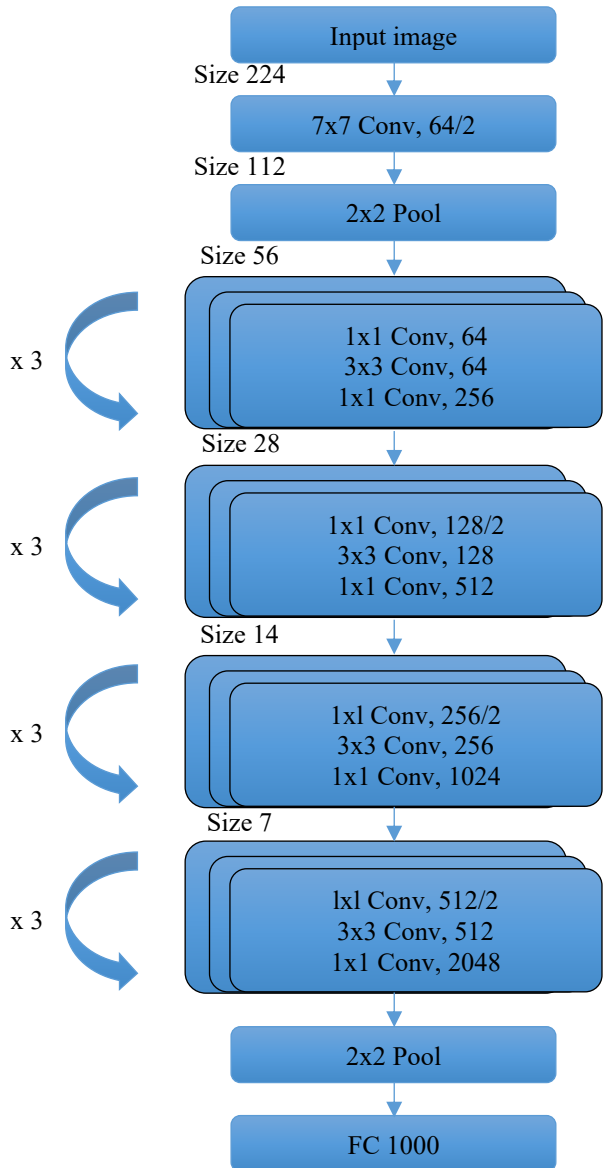

**Fig. 5 An in-depth analysis of the dimensions and convolutional layers of ResNet-50**

Figure 5 gives a thorough perspective of the ResNet-50 architecture, as seen in the image. An input image of size 224 × 224 is first processed using a 7 x 7 convolutional layer with 64 filters, and then max pooling is applied. The network is divided into four phases, with several residual blocks of 1x1, 3x3, and 1x1 convolutions in each stage. The image's spatial dimensions decrease from 112 by 112 to 7 by 7 as it moves through these steps, and the number of filters rises. The mathematics that goes behind the Multiple convolutional layers, batch normalization, and ReLU can be elaborated in the following steps:

*3.2.1. Convolution Operation*
In the ResNet-50 architecture, the convolution operation is a key procedure that acts as the main feature extraction technique. This process entails swiping a filter, sometimes called a kernel, to create an activation map over the input image. Typically, the filter is a *k x k* matrix. The stride controls how much the filter moves throughout the image. *(a - k + 2p)/s + 1 X (b - k + 2p)/s + 1* is the output size obtained by applying convolution to an input picture of dimensions *a x b* with a kernel of size *k*, padding *p*, and stride *s* [30]. Convolutional layers are essential to ResNet-50's ability to recognize and learn distinct characteristics at different abstraction levels.

The convolution operation's equation is provided by:

$$(I * K)(i.j) = \sum_m \sum_n I(i + m, j + n) K(m.n) \quad (1)$$

Where *I(i,j)* is the input image, and *K(m,n)* represents the convolutional filter.

*3.2.2. Residual Block Equation*
The essential component of ResNet-50 that allows the network to train deep architectures effectively is the residual block. A residual block is made up of multiple convolutional layers. The input $x$ is then added directly to the output of these layers, forming a "skip connection" or "shortcut." Thanks to this approach, the network can learn residual functions-which are simpler to optimize. One formulates the output of a residual block as follows:

$$Output = F(x) + x \quad (2)$$

The function that the convolutional layers inside the block have learnt is represented by $(x)$ in Equation (2), where $x$ is the block's input. This addition facilitates the training of deeper networks by mitigating the degradation issue.

*3.2.3. Activation Function (ReLU)*
The activation function that is employed throughout the ResNet-50 architecture is the Rectified Linear Unit (ReLU). It gives the network non-linearity, which is necessary for figuring out intricate patterns. If the input is positive, the ReLU function outputs it directly; if not, it produces zero, retaining only the positive values. The definition of the ReLU function is:

$$ReLU(x) = max(0, x) \quad (3)$$

Applying nonlinear modifications to the input at each layer enables the network to learn from the data, and this straightforward but powerful function is essential to this process.

*3.2.4. Batch Normalization*
ResNet-50 uses batch normalization to normalize activations in every network layer. This procedure ensures that the input distribution to each layer is constant during training, which helps stabilize the learning process.

The Equation (4) describes the normalizing process:

$$\hat{x} = \frac{x - \mu}{\sqrt{\sigma^2 + \epsilon}} \quad (4)$$

Where $\mu$ and $\sigma^2$ are the mean and variance of activations, and $\epsilon$ has been introduced to prevent division by zero. This method aids in accelerating training and enhancing model performance.

*3.2.5. Softmax Function*
For classification tasks, the last layer of ResNet-50 employs the Softmax function. It transforms the network's raw output values, or logits, into probabilities that are easier to categorise. The function ensures that the values it outputs sum up to 1, which means you can use them to calculate the probability for each class. The definition of the Softmax function is:

$$Softmax(Z_i) = \frac{e^{z_i}}{\sum_j e^{z_j}} \quad (5)$$

In Equation (5), $z_i$ Represents the logit corresponding to class $i$, and the total of each logit's exponentials is the denominator. By doing this, the normalization of each output value and the probability of each class are guaranteed. The ResNet-50 architecture is a deep learning model that excels in WBC categorization due to its residual connections and deep learning capabilities [7]. It uses convolutional layers to extract characteristics and patterns from WBC images, allowing it to learn from complex and diverse characteristics efficiently. This is particularly useful for differentiating between different WBC types, such as neutrophils, monocytes, and lymphocytes. ResNet-50 can reliably categorize WBC pictures based on morphological features and is well-suited for medical imaging jobs where accurate and reliable classification is crucial.

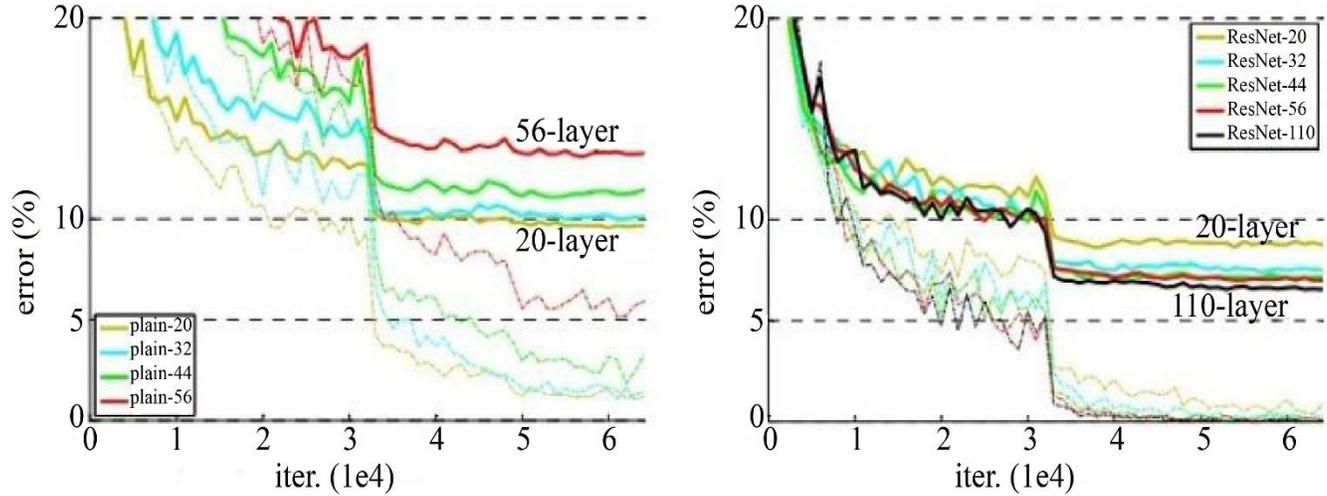

**Fig. 6 CNNs versus ResNets: depth's effect on accuracy [43]**

The image in Figure 6 contrasts ResNet models and conventional CNNs, showing that deeper simple networks have higher error rates. ResNet models lower errors and use residual connections to reduce vanishing gradient issues.

*3.3. Model Used for Image Enhancement*
An improved ESRGAN model called Real Enhanced Super-Resolution Generative Adversarial Network (Real-ESRGAN) was created to address real-world image super-resolution problems. Real-ESRGAN is designed to manage the complex and diverse degradation found in real-world photos, in contrast to the original ESRGAN, which was primarily focused on providing visually pleasing results from relatively clean images. This improvement is made possible by introducing a high-order degradation model, which produces more reliable and broadly applicable super-resolution outputs by simulating several real-world degradations like blur, noise, and compression artifacts. For applications where preserving the integrity of minute features is crucial, such as medical imaging, the advancements gained in Real-ESRGAN are quite helpful [17].

To enhance its performance on real-world photos, Real-ESRGAN expands upon the fundamental architecture of ESRGAN and makes several significant changes. The Residual-in-Residual Dense Block (RRDB) architecture used by the generator in Real-ESRGAN enables it to capture more complicated data and provide outputs of greater quality. Another noteworthy improvement is the spectral normalization-based U-Net-based discriminator, which offers more thorough per-pixel feedback and improves the model's ability to handle a variety of degradations. These architectural modifications guarantee that Real-ESRGAN can efficiently upscale photos while maintaining important features, which makes it a valuable tool in situations where precise analysis requires high-resolution images.

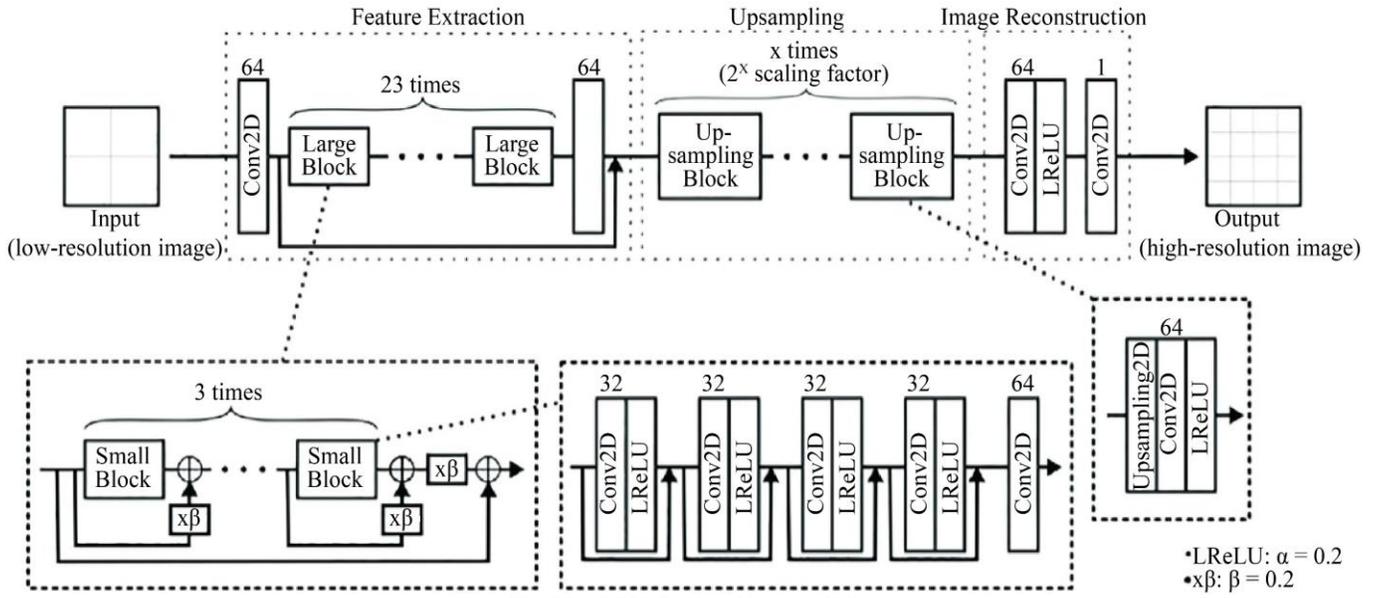

**Fig. 7 Real-ESRGAN architecture [44]**

Figure 7 shows that the Real-ESRGAN uses a generator network similar to ESRGAN [17]. Real-ESRGAN is trained by mimicking realistic image degradations to build an extensive training dataset. The high-resolution photos in this collection are subjected to many layers of blur, noise, downsampling, and compression, thanks to a customizable degradation model. This enables the generator to learn how to recover images from various degraded inputs efficiently. Pixel loss, perceptual loss, and GAN loss are among the losses that serve as training guidelines.

These losses work in concert to guarantee that the generated images are both highly resolved and perceptually close to the ground truth. The upscaled images are more realistic and detailed due mainly to the perceptual loss, which is calculated using a pre-trained VGG network.

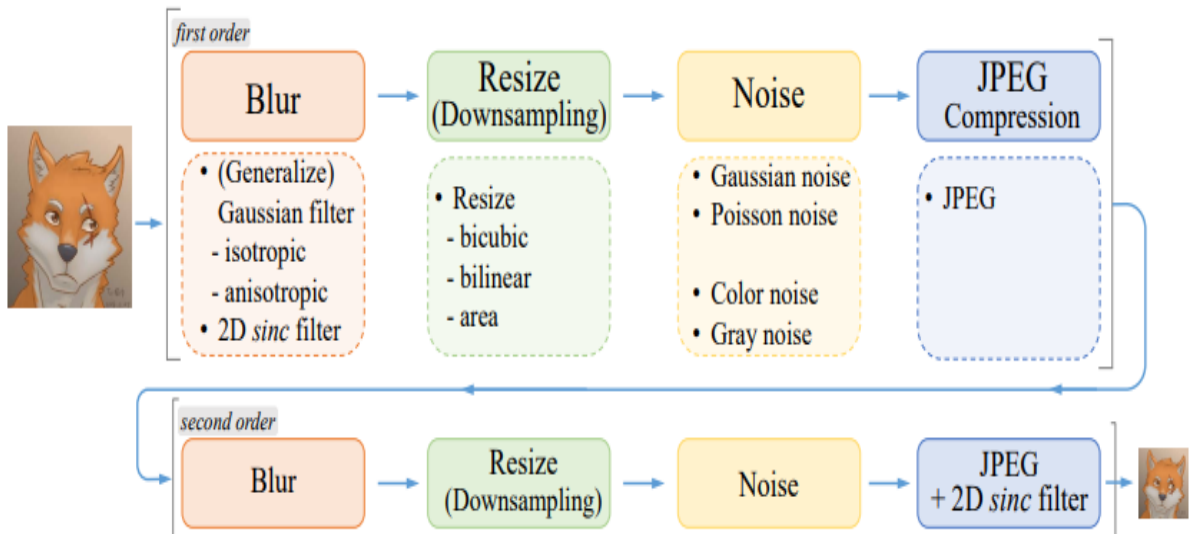

**Fig. 8 Training process of real-ESRGAN**

Figure 8 provides an overview of Real-ESRGAN's use of pure synthetic data generation. In order to simulate more realistic degradations, it uses a second-order deterioration process, in which each degradation process applies the classical degradation model. There is a complete list of options for JPEG compression, noise reduction, resizing, and blur. Additionally, we use a sinc filter to create typical overshoot and ringing effects [17]. Real-ESRGAN was utilized in the study to improve the resolution of WBC pictures, focusing on the problems related to the ResNet-50 model's inability to accurately categorize neutrophil and eosinophil cells [17]. Real-ESRGAN assisted in enhancing the data quality by upscaling the training pictures, giving the model access to more intricate features for learning. This improvement is especially significant for medical imaging, where minute variations in cell types might be critical to precise classification [19]. The model could now more accurately and reliably identify the various types of WBC thanks to the enhanced resolution and detail provided by Real-ESRGAN, which probably also helped to reduce classification mistakes [22].

## 4. Results and Discussions
The study aimed to enhance WBC images using Real-ESRGAN for image enhancement before feeding into a ResNet-50 model. However, testing revealed a slight decrease in accuracy, possibly due to artifacts or altered features. The findings suggest that while image enhancement techniques can improve visual quality, they may not necessarily improve classification performance.

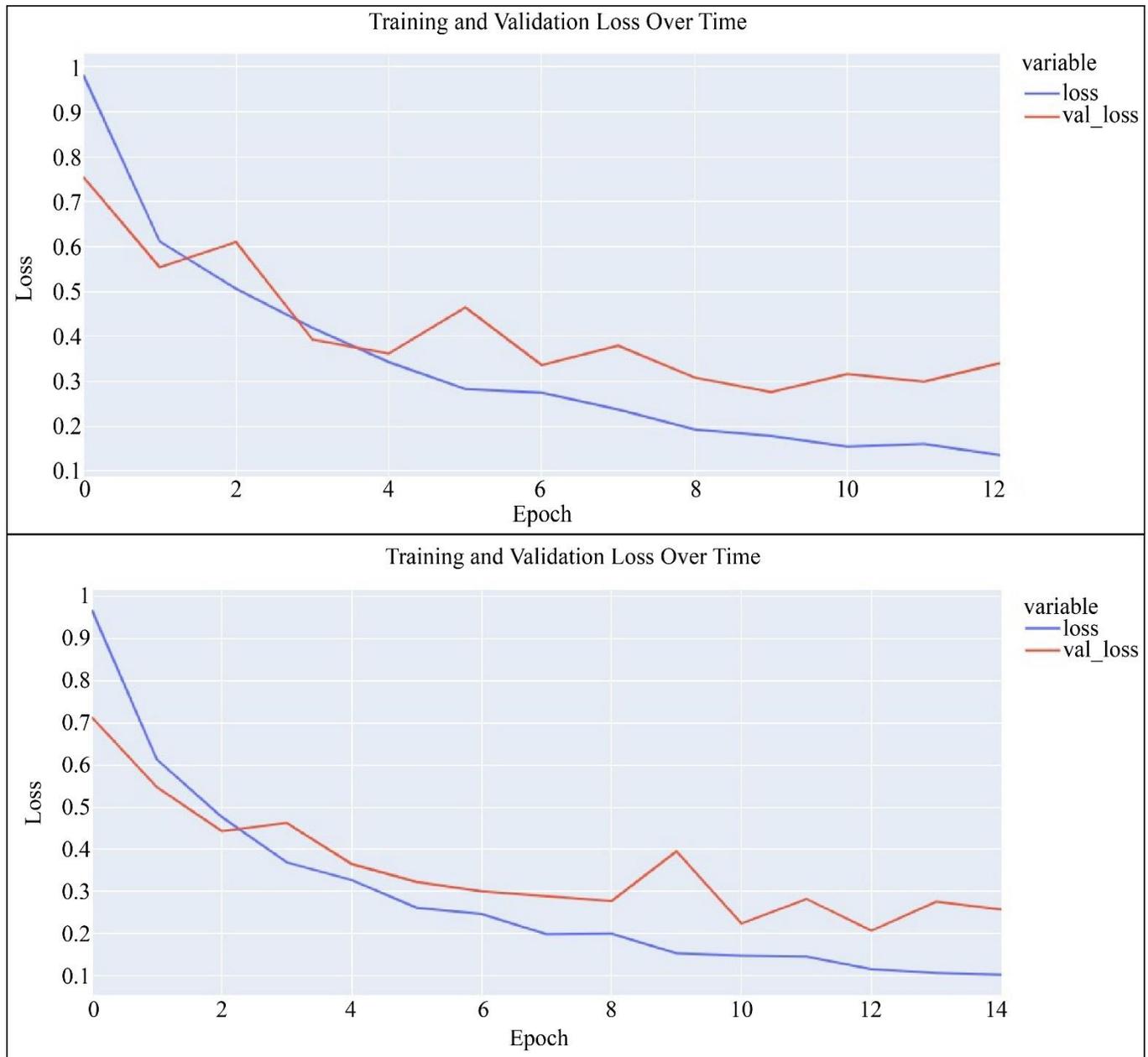

**Fig. 9 Comparison of training and validation over time**

In Figure 9, the classification accuracy of ResNet-50 was not significantly improved by using Real-ESRGAN to enhance images of WBCs. The training and validation performance was comparable to the original photos, indicating that the higher resolution was not helpful for this particular task.

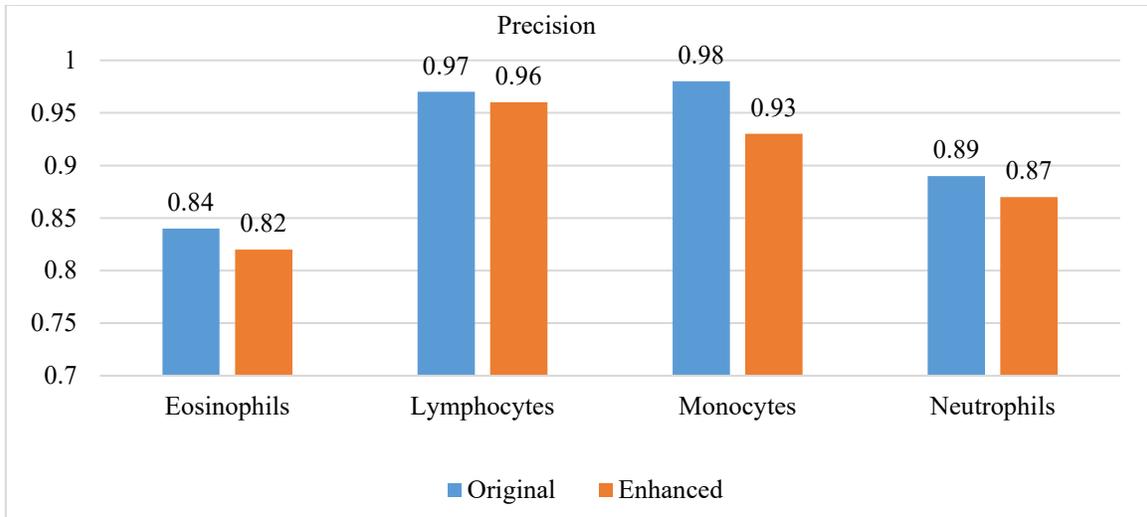
**Fig. 10 Comparison of precision**

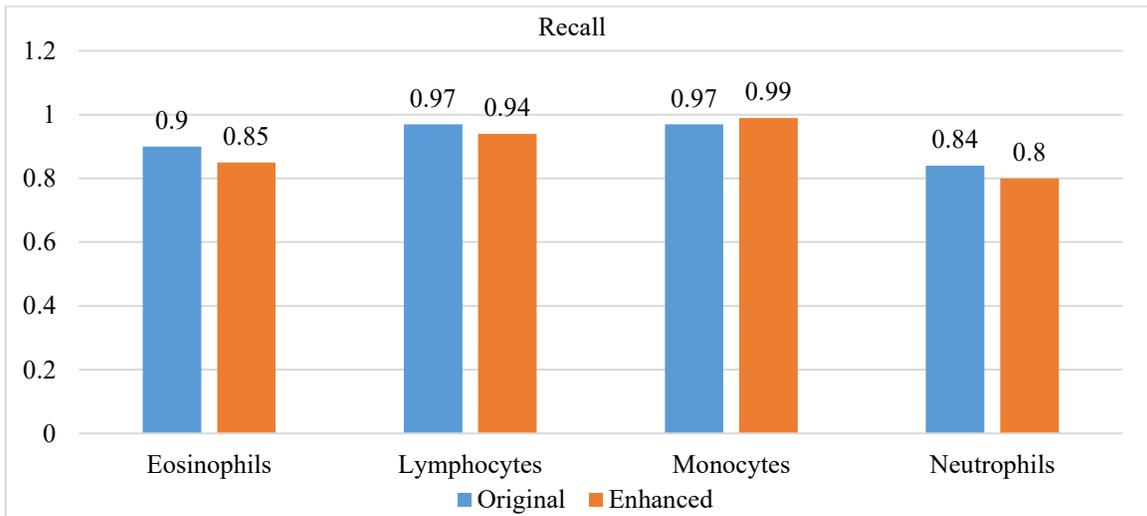
**Fig. 11 Comparison of recall**

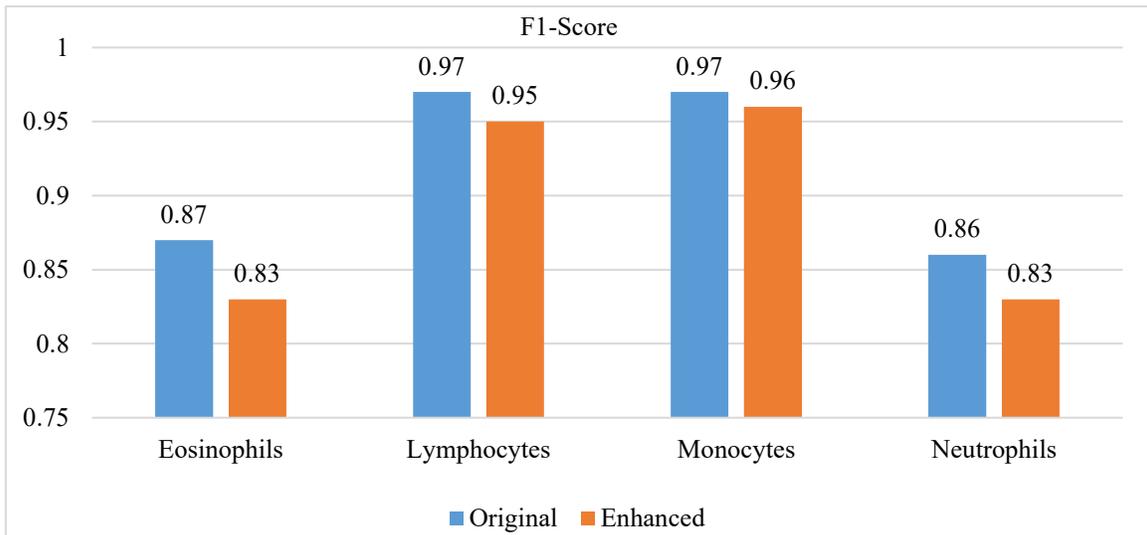
**Fig. 12 Comparison of F1-score**

Bar graphs for accuracy, recall, and F1-Score were used to evaluate the performance of ResNet-50 on the original and improved images in Figures 10, 11, and 12. The old dataset consistently outperformed the upgraded dataset by a small margin in every statistic. The enhanced pictures led to a modest drop in the model's overall accuracy, recall, and F1-Score, particularly for these two cell types.

Table 3. Classification report of ResNet-50 on an original dataset

| Class | Precision | Recall | F1-Score | Support |
|---|---|---|---|---|
| Eosinophils | 0.84 | 0.90 | 0.87 | 499 |
| Lymphocytes | 0.97 | 0.97 | 0.97 | 496 |
| Monocytes | 0.98 | 0.97 | 0.97 | 495 |
| Neutrophils | 0.89 | 0.84 | 0.860 | 499 |
| Accuracy | - | - | 0.92 | 1989 |
| Macro- avg | 0.92 | 0.92 | 0.92 | 1989 |
| Weighted-avg | 0.92 | 0.92 | 0.92 | 1989 |

Table 4. Classification report of ResNet-50 on an enhanced dataset

| Class | Precision | Recall | F1-Score | Support |
|---|---|---|---|---|
| Eosinophils | 0.82 | 0.85 | 0.83 | 499 |
| Lymphocytes | 0.96 | 0.94 | 0.95 | 496 |
| Monocytes | 0.93 | 0.99 | 0.96 | 495 |
| Neutrophils | 0.87 | 0.80 | 0.83 | 499 |
| Accuracy | - | - | 0.89 | 1989 |
| Macro- avg | 0.89 | 0.89 | 0.89 | 1989 |
| Weighted-avg | 0.89 | 0.89 | 0.89 | 1989 |

Even then, monocytes displayed a little increase in recollection after using the improved photos. The graphs emphasize the trade-offs associated with using image improvement algorithms like Real-ESRGAN, showing that although enhancement may positively affect some categories, it may negatively impact others.

Tables 3 and 4 summarise classification performances on original and improved datasets. Both the tables act as a substantiation of the findings. The model's overall accuracy with the original dataset was 92%, with excellent precision, recall, and F1 scores for all cell types. Neutrophils and Eosinophils scored marginally lower than those of the other classes.

Applied to the improved dataset, the accuracy fell to 89% overall. The results showed that although monocytes performed somewhat better, the precision and recall of neutrophils and eosinophils decreased. This suggests that not all cell types benefited equally from the image enhancement, with neutrophils and eosinophils significantly affected.

Tables 5 and 6 show the information on testing the model on both types of images. The influence of improving the picture dataset on model performance is not visible when comparing the two confusion matrices. Compared to the original dataset, there is not any improvement in classification.

In contrast, we can see a slight deterioration in the model's performance, especially in classifying the images of Eosinophils and Neutrophils. The improved dataset exhibits a deterioration in correctly classifying the data throughout the validation dataset, suggesting that the model is more adept at discriminating. This implies that picture enhancement may not have added consistency to the model's categorization and did not significantly improve.

Table 5. Confusion matrix of the model tested on the original dataset

| Actual \ Predicted | Eosinophils | Lymphocytes | Monocytes | Neutrophils |
|---|---|---|---|---|
| Eosinophils | 447 | 9 | 0 | 43 |
| Lymphocytes | 9 | 481 | 0 | 6 |
| Monocytes | 10 | 1 | 479 | 5 |
| Neutrophils | 65 | 6 | 10 | 418 |

Table 6. Confusion matrix of the model tested on the enhanced images dataset

| Actual \ Predicted | Eosinophils | Lymphocytes | Monocytes | Neutrophils |
|---|---|---|---|---|
| Eosinophils | 423 | 9 | 12 | 55 |
| Lymphocytes | 18 | 466 | 5 | 7 |
| Monocytes | 5 | 2 | 488 | 0 |
| Neutrophils | 73 | 6 | 20 | 400 |

## 5. Conclusion

In this work, a detailed investigation has been carried out on how the ResNet-50 model's ability to classify WBC pictures was affected by using Real-ESRGAN for image enhancement. The main objective was to ascertain whether improving the photos' resolution and quality may result in better feature extraction and, in turn, better classification performance. The outcome, meanwhile, fell short of our expectations. Using Real-ESRGAN did not improve the model's performance; on the contrary, it caused either no discernible change in accuracy or a little reduction in it. This surprising result implies that although Real-ESRGAN is good at improving pictures visually, it could add artifacts or change significant characteristics in a way that makes the model less successful at classifying the photos. The results emphasize how intricate deep learning pipeline preprocessing methods may be. Improving picture quality does not always result in a better model; this is especially true for classification jobs where the finer details of the image data are crucial. These findings emphasize the necessity of carefully assessing preprocessing techniques in light of the intended use. Prospective research endeavours may investigate substitute methods for augmenting images or modifying model structures to more effectively include refined datasets, guaranteeing that aesthetic enhancement in data favourably impacts categorization precision.

## References

[1] Ekta Gavas, and Kaustubh Olpadkar, "Deep CNNs for Peripheral Blood Cell Classification," *arXiv Preprint*, 2021. [CrossRef] [Google Scholar] [Publisher Link]

[2] Qasem Abu Al-Haija, and Adeola Adebanjo, "Breast Cancer Diagnosis in Histopathological Images Using Resnet-50 Convolutional Neural Network," *2020 IEEE International IOT, Electronics and Mechatronics Conference (IEMTRONICS)*, Vancouver, BC, Canada, pp. 1-7, 2020. [CrossRef] [Google Scholar] [Publisher Link]

[3] Hua Chen et al., "Accurate Classification of White Blood Cells by Coupling Pre-Trained Resnet and Densenet with SCAM Mechanism," *BMC Bioinformatics*, vol. 23, no. 1, pp. 1-20, 2022. [CrossRef] [Google Scholar] [Publisher Link]

[4] Samit Shivadekar et al., "Deep Learning Based Image Classification of Lungs Radiography for Detecting COVID-19 Using A Deep CNN And Resnet 50," *International Journal of Intelligent Systems and Applications in Engineering*, vol. 11, no. 1S, pp. 241-250, 2023. [Google Scholar] [Publisher Link]

[5] Thinam Tamang, Sushish Baral, and May Phu Paing, "Classification of White Blood Cells: A Comprehensive Study Using Transfer Learning Based on Convolutional Neural Networks," *Diagnostics*, vol. 12, no. 12, pp. 1-12, 2022. [CrossRef] [Google Scholar] [Publisher Link]

[6] Fatima-Zohra Hamlili et al., "Transfer Learning with Resnet-50 for Detecting COVID-19 in Chest X-Ray Images," *Indonesian Journal of Electrical Engineering and Computer Science*, vol. 25, no. 3, pp. 1458-1468, 2022. [CrossRef] [Google Scholar] [Publisher Link]

[7] Jawad Yousef Ibrahim Alzamily, Syaiba Balqish Ariffin, and Samy S. Abu Naser, "Classification of Encrypted Images Using Deep Learning-Resnet50," *Journal of Theoretical and Applied Information Technology*, vol. 100, no. 21, pp. 6610-6620, 2022. [Google Scholar] [Publisher Link]

[8] Priya Aggarwal et al., "COVID-19 Image Classification Using Deep Learning: Advances, Challenges and Opportunities," *Computers in Biology and Medicine*, vol. 144, 2022. [CrossRef] [Google Scholar] [Publisher Link]

[9] Raj Kumar et al., "Evaluation of Deep Learning based Resnet-50 for Plant Disease Classification with Stability Analysis," *2022 6th International Conference on Intelligent Computing and Control Systems (ICICCS)*, Madurai, India, pp. 1280-1287, 2022. [CrossRef] [Google Scholar] [Publisher Link]

[10] Chun-Ling Lin, and Kun-Chi Wu, "Development of Revised ResNet-50 for Diabetic Retinopathy Detection," *BMC Bioinformatics*, vol. 24, no. 1, pp. 1-18, 2023. [CrossRef] [Google Scholar] [Publisher Link]

[11] Rudresh Pillai et al., "Brain Tumor Classification Using VGG 16, Resnet50, and Inception V3 Transfer Learning Models," *2023 2nd International Conference for Innovation in Technology (INOCON)*, Bangalore, India, pp. 1-5, 2023. [CrossRef] [Google Scholar] [Publisher Link]

[12] Mohammad Zolfaghari, and Hedieh Sajedi, "A Survey on Automated Detection and Classification of Acute Leukemia and Wbcs in Microscopic Blood Cells," *Multimedia Tools and Applications*, vol. 81, no. 5, pp. 6723-6753, 2022. [CrossRef] [Google Scholar] [Publisher Link]

[13] Amin Khouani et al., "Automated Recognition of White Blood Cells Using Deep Learning," *Biomedical Engineering Letters*, vol. 10, no. 3, pp. 359-367, 2020. [CrossRef] [Google Scholar] [Publisher Link]

[14] Areej Malkawi et al., "White Blood Cells Classification Using Convolutional Neural Network Hybrid System," *2020 IEEE 5th Middle East and Africa Conference on Biomedical Engineering (MECBME)*, Amman, Jordan, pp. 1-5, 2020. [CrossRef] [Google Scholar] [Publisher Link]

[15] Jingan Liu, and Naiwala P. Chandrasiri, "CA-ESRGAN: Super-Resolution Image Synthesis Using Channel Attention-Based ESRGAN," *IEEE Access*, vol. 12, pp. 25740-25748, 2024. [CrossRef] [Google Scholar] [Publisher Link]

[16] Zhengwei Zhu et al., "IRE: Improved Image Super-Resolution Based on Real-ESRGAN," *IEEE Access*, vol. 11, pp. 45334-45348, 2023. [CrossRef] [Google Scholar] [Publisher Link]

[17] Xintao Wang et al., "Real-ESRGAN: Training Real-World Blind Super-Resolution with Pure Synthetic Data," *2021 IEEE/CVF International Conference on Computer Vision Workshops (ICCVW)*, Montreal, BC, Canada, pp. 1905-1914, 2021. [CrossRef] [Google Scholar] [Publisher Link]

[18] Baptiste Roziere et al., "Tarsier: Evolving Noise Injection in Super-Resolution Gans," *2020 25th International Conference on Pattern Recognition (ICPR)*, Milan, Italy, pp. 7028-7035, 2021. [CrossRef] [Google Scholar] [Publisher Link]

[19] Zihao Wei et al., "A-ESRGAN: Training Real-World Blind Super-Resolution with Attention U-Net Discriminators," *Pacific Rim International Conference on Artificial Intelligence*, Jakarta, Indonesia, pp. 16-27, 2023. [CrossRef] [Google Scholar] [Publisher Link]

[20] Haoyu Ren et al., "Real-World Super-Resolution Using Generative Adversarial Networks," *2020 IEEE/CVF Conference on Computer Vision and Pattern Recognition Workshops (CVPRW)*, pp. 1760-1768, 2020. [CrossRef] [Google Scholar] [Publisher Link]

[21] Rao Muhammad Umer, Rao Muhammad Umer, and Christian Micheloni, "Deep Generative Adversarial Residual Convolutional Networks for Real-World Super-Resolution," *2020 IEEE/CVF Conference on Computer Vision and Pattern Recognition Workshops (CVPRW)*, Seattle, WA, USA, pp. 438-439, 2020. [CrossRef] [Google Scholar] [Publisher Link]

[22] Campbell D. Watson et al., "Investigating Two Super-Resolution Methods for Downscaling Precipitation: ESRGAN and CAR," *arXiv Preprint*, 2020. [CrossRef] [Google Scholar] [Publisher Link]

[23] Jie Song et al., "Dual Perceptual Loss for Single Image Super-Resolution Using ESRGAN," *arXiv Preprint*, 2022. [CrossRef] [Google Scholar] [Publisher Link]

[24] Yuval Bahat, and Tomer Michaeli, "Explorable Super Resolution," *2020 IEEE/CVF Conference on Computer Vision and Pattern Recognition (CVPR)*, Seattle, WA, USA, pp. 2713-2722, 2020. [CrossRef] [Google Scholar] [Publisher Link]

[25] Zewen Li et al., "A Survey of Convolutional Neural Networks: Analysis, Applications, and Prospects," *IEEE Transactions on Neural Networks and Learning Systems*, vol. 33, no. 12, pp. 6999-7019, 2022. [CrossRef] [Google Scholar] [Publisher Link]

[26] Tusneem A. Elhassan et al., "Classification of Atypical White Blood Cells in Acute Myeloid Leukemia Using a Two-Stage Hybrid Model Based on Deep Convolutional Autoencoder and Deep Convolutional Neural Network," *Diagnostics*, vol. 13, no. 2, pp. 1-20, 2023. [CrossRef] [Google Scholar] [Publisher Link]

[27] Hüseyin Kutlu, Engin Avci, and Fatih Özyurt, "White Blood Cells Detection and Classification Based on Regional Convolutional Neural Networks," *Medical Hypotheses*, vol. 135, 2020. [CrossRef] [Google Scholar] [Publisher Link]

[28] Kitsuchart Pasupa, Supawit Vatathanavaro, and Suchat Tungjitnob, "Convolutional Neural Networks Based Focal Loss for Class Imbalance Problem: A Case Study of Canine Red Blood Cells Morphology Classification," *Journal of Ambient Intelligence and Humanized Computing*, vol. 14, no. 11, pp. 15259-15275, 2020. [CrossRef] [Google Scholar] [Publisher Link]

[29] Madona B. Sahaai et al., "Resnet-50 Based Deep Neural Network Using Transfer Learning for Brain Tumor Classification," *AIP Conference Proceedings*, vol. 2463, no. 1, 2022. [CrossRef] [Google Scholar] [Publisher Link]

[30] Deepika Kumar et al., "Automatic Detection of White Blood Cancer from Bone Marrow Microscopic Images Using Convolutional Neural Networks," *IEEE Access*, vol. 8, pp. 142521-142531, 2020. [CrossRef] [Google Scholar] [Publisher Link]

[31] Sarang Sharma et al., "[Retracted] Deep Learning Model for the Automatic Classification of White Blood Cells," *Computational Intelligence and Neuroscience*, vol. 2022, pp. 1-13, 2022. [CrossRef] [Google Scholar] [Publisher Link]

[32] Li Ma et al., "Combining DC-GAN with ResNet for Blood Cell Image Classification," *Medical & Biological Engineering & Computing*, vol. 58, pp. 1251-1264, 2020. [CrossRef] [Google Scholar] [Publisher Link]

[33] Guozhen Chen et al., "Prediction of Chronic Kidney Disease Using Adaptive Hybridized Deep Convolutional Neural Network on the Internet of Medical Things Platform," *IEEE Access*, vol. 8, pp. 100497-100508, 2020. [CrossRef] [Google Scholar] [Publisher Link]

[34] Aniruddha Dutta et al., "An Efficient Convolutional Neural Network for Coronary Heart Disease Prediction," *Expert Systems with Applications*, vol. 159, 2020. [CrossRef] [Google Scholar] [Publisher Link]

[35] D.R. Sarvamangala, and Raghavendra V. Kulkarni, "Convolutional Neural Networks in Medical Image Understanding: A Survey," *Evolutionary Intelligence*, vol. 15, no. 1, pp. 1-22, 2022. [CrossRef] [Google Scholar] [Publisher Link]

[36] Grace W. Lindsay, "Convolutional Neural Networks as a Model of the Visual System: Past, Present, and Future," *Journal of Cognitive Neuroscience*, vol. 33, no. 10, pp. 2017-2031, 2021. [CrossRef] [Google Scholar] [Publisher Link]

[37] Saeed Iqbal et al., "On the Analyses of Medical Images Using Traditional Machine Learning Techniques and Convolutional Neural Networks," *Archives of Computational Methods in Engineering*, vol. 30, no. 5, pp. 3173-3233, 2023. [CrossRef] [Google Scholar] [Publisher Link]

[38] Elima Hussain et al., "A Comprehensive Study on the Multi-Class Cervical Cancer Diagnostic Prediction on Pap Smear Images Using a Fusion-Based Decision from Ensemble Deep Convolutional Neural Network," *Tissue and Cell*, vol. 65, 2020. [CrossRef] [Google Scholar] [Publisher Link]

[39] Youhui Tian, "Artificial Intelligence Image Recognition Method Based on Convolutional Neural Network Algorithm," *IEEE Access*, vol. 8, pp. 125731-125744, 2020. [CrossRef] [Google Scholar] [Publisher Link]

[40] Teja Kattenborn et al., "Review on Convolutional Neural Networks (CNN) in Vegetation Remote Sensing," *ISPRS Journal of Photogrammetry and Remote Sensing*, vol. 173, pp. 24-49, 2021. [CrossRef] [Google Scholar] [Publisher Link]

[41] Ademola E. Ilesanmi, and Taiwo O. Ilesanmi, "Methods for Image Denoising Using Convolutional Neural Network: A Review," *Complex & Intelligent Systems*, vol. 7, no. 5, pp. 2179-2198, 2021. [CrossRef] [Google Scholar] [Publisher Link]

[42] Ananda S. Paymode, and Vandana B. Malode, "Transfer Learning for Multi-Crop Leaf Disease Image Classification Using Convolutional Neural Network VGG," *Artificial Intelligence in Agriculture*, vol. 6, pp. 23-33, 2022. [CrossRef] [Google Scholar] [Publisher Link]

[43] Kaiming He et al., "Deep Residual Learning for Image Recognition," *2016 IEEE Conference on Computer Vision and Pattern Recognition (CVPR)*, Las Vegas, NV, USA, pp. 770-778, 2016. [CrossRef] [Google Scholar] [Publisher Link]

[44] Motoharu Sonogashira, Michihiro Shonai, and Masaaki Iiyama, "High-Resolution Bathymetry by Deep-Learning-Based Image Superresolution," *PLOS One*, vol. 15, no. 7, pp. 1-19, 2020. [CrossRef] [Google Scholar] [Publisher Link]